\documentclass[aps,prl,showpacs,reprint,floatfix]{revtex4-1}
\usepackage{bm}
\usepackage[dvips]{graphicx}
\usepackage[utf8]{inputenc}
\usepackage[T1]{fontenc}
\usepackage{color}

\begin{document}

\title{Band filling control of the Dzyaloshinskii-Moriya interaction in weakly ferromagnetic insulators}

\author{G. Beutier$^{1}$, S. P. Collins$^{2}$, O. V. Dimitrova$^{8}$, V. E. Dmitrienko$^{3}$, M.I. Katsnelson$^{4,5}$, Y.O. Kvashnin$^{6}$, A.I. Lichtenstein$^{5,7}$, V. V. Mazurenko$^{5}$, G. Nisbet$^{2}$,  E. N. Ovchinnikova$^{8}$, D. Pincini$^{2,9}$}

\affiliation{
$^{1}$ Univ. Grenoble Alpes, CNRS, Grenoble INP\thanks{Institute of Engineering Univ. Grenoble Alpes}, SIMaP, F-38000 Grenoble, France\\
$^{2}$ Diamond Light Source Ltd, Diamond House, Harwell Science and Innovation Campus, Didcot, Oxfordshire, OX11 0DE, United Kingdom \\
$^{3}$ A. V. Shubnikov Institute of Crystallography RAS, Moscow 119333, Russia \\
$^{4}$ Radboud University Nijmegen, Institute for Molecules and Materials, Heyendaalseweg 135, NL-6525 AJ Nijmegen, The Netherlands \\
$^{5}$ Department of Theoretical Physics and Applied Mathematics, Ural Federal University, Mira str. 19, 620002 Ekaterinburg, Russia \\
$^{6}$ Department of Physics and Astronomy, Division of Materials Theory, Uppsala University, Box 516, SE-75120 Uppsala, Sweden \\
$^{7}$ I. Institut f{\"u}r Theoretische Physik, Universit{\"a}t Hamburg, Jungiusstra{\ss}e 9, D-20355 Hamburg, Germany \\
$^{8}$ M.V.Lomonosov Moscow State University, Leninskie Gory, Moscow 119991, Russia \\
$^{9}$ London Centre for Nanotechnology and Department of Physics and Astronomy, University College London, London WC1E 6BT, United Kingdom
}

\begin{abstract}
We observe and explain theoretically a dramatic evolution of the Dzyaloshinskii-Moriya interaction in the series
of isostructural weak ferromagnets, MnCO$_3$, FeBO$_3$, CoCO$_3$ and NiCO$_3$. The sign of the interaction is encoded in the phase of the x-ray magnetic diffraction amplitude, 
observed through interference with resonant quadrupole scattering. We find very good quantitative agreement with first-principles electronic structure
calculations, reproducing both sign and magnitude through the series, and propose a simplified `toy model' to explain the change in sign with 3$d$ shell filling.
The model gives insight into the evolution of the DMI in Mott and charge transfer insulators.
\end{abstract}


\maketitle

{\it Introduction.} 
The Dzyaloshinskii-Moriya interaction (DMI) appears in magnetic materials with, at least locally, broken inversion symmetry. 
It leads to an exchange energy that scales with the vector product of spins
$\bm {S_1} \times \bm {S_2}$ 
and is thus antisymmetric with respect to interchange of the spins, favouring non-collinear order. 
First introduced to explain the canting of moments in weak ferromagnets \cite{Dzyaloshinsky}, with a microscopic origin in spin-orbit coupling (SOC) \cite{Moriya1,Moriya2}, the DMI has recently been shown
to be responsible for the stabilization of various exotic non-collinear magnetic ground-states, such as spin-spirals \cite{bode} and skyrmions \cite{skyrm1,skyrm2,skyrm3}. 
Such magnetic orders are of particular interest from both fundamental and applied points of view.
For instance, skyrmions are topologically protected states, which makes them promising for novel spintronic applications. 
DMI is an important ingredient in multiferroics with spiral magnetic order, where it is thought to promote an electric polarisation either by polarizing electronic orbitals \cite{Katsura} or by inducing atomic displacements \cite{Sergienko}.
DMI stabilizes chiral domain walls, which can be driven by current rather than magnetic field \cite{emori, PhysRevLett.116.147204}.
Also, they can be used for manipulation of spin wave currents ("magnon transistor") \cite{PhysRevLett.116.147204}.
The possibility to control and change the sign of the DMI in magnetic materials is an essential step towards finding suitable materials for spintronics applications.
Up to date such manipulation has been experimentally realized for the isostructural $B20$ metallic alloys  Fe$_{1-x}$Co$_{x}$Si \cite{FeCoSi},  
Mn$_{1-x}$Fe$_{x}$Ge \cite{MnFeGe} and  Fe$_{1-x}$Co$_{x}$Ge \cite{FeCoGe} demonstrating a very complex and rich magnetic phase diagram depending on the doping and the applied magnetic field.

The magnitude of the DMI has been evaluated in several weak ferromagnets where it is related to the magnitude of the canting angle and thus to the net magnetization.
Its sign, however, has been determined experimentally in only a handful of such materials \cite{MoriyaNiF2,Shulman,BrownNiF2,MoskvinFeF3, BrownGd2CuO4, DmitrienkoNP, Fe2O3, YVO4}. 
Moreover, these compounds have different crystal structures, rendering any systematic trends meaningless. 

Here we report a systematic experimental and theoretical study of the insulator-counterpart of the systems with tunable DMI: 
isostructural MnCO$_{3}$, FeBO$_3$, CoCO$_3$ and NiCO$_3$, with $R\bar{3}c$ crystal symmetry.
In contrast to the metallic $B20$ alloys with competing long-range magnetic interactions, strongly affected by the dynamical Coulomb correlations \cite{Mazurenko},
the magnetic structure of these $R\bar{3}c$ insulators is much simpler.
In these systems, every metal atom interacts predominantly with its six nearest neighbors, 
providing a route to a truly microscopic understanding of the DMI.

\begin{figure}[!ht]
 \includegraphics[width=\columnwidth]{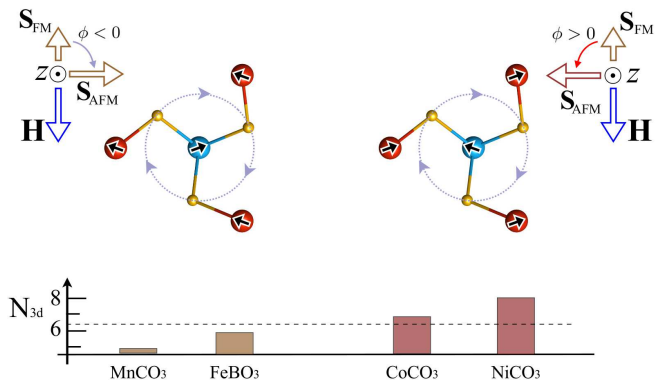}
 \caption{Local atomic and magnetic orders in the weak ferromagnets of this work. 
 The ions of the two magnetic sublattices are represented by blue and red spheres, with black arrows denoting the direction of their spins.
 Oxygen atoms between the two adjacent transition metal layers are represented as yellow spheres. The dotted circles highlight the twist of the oxygen layer. The bottom panel shows the occupation of the 3d level of a magnetic ion. The left and right panels show the two possible magnetic configurations which stabilize depending on the $3d$ occupation and, therefore, the sign of the DMI, for a net ferromagnetic moment pointing along the magnetic field $\mathbf{H}$. 
 $\mathbf{S}_{AFM}$ denotes the direction of the antiferromagnetic spin structure.}
 \label{FIG1}
\end{figure}

The four crystals studied here have the same crystal structure \cite{Wyckoff,Diehl,Maslen,Pertlik}, 
consisting in a stack of alternating $3d$ transition metal (TM) and oxygen/carbon (oxygen/boron) layers (Fig.~\ref{FIG1}). 
The TM ions occupy the centre of elongated \textit{M}O$_6$ octahedra ($M\in$ \{Mn, Fe, Co, Ni\}). 
The exchange interaction between the TM ions is mediated by the oxygen ions.
The structural twist of the oxygen layers with respect to the TM layers (Fig.~\ref{FIG1})
shifts the oxygen atoms away from the middle point between TM atoms and 
breaks the inversion symmetry at the oxygen sites, thus allowing the DMI interaction between the TM sites.
This twist alternates in sign from one oxygen layer to the next, such that the crystal is globally centrosymmetric.

These crystals have the same antiferromagnetic order, if one ignores the small ferromagnetic component: 
the magnetic moments are in the basal plane, aligned parallel in a single TM layer and antiparallel between adjacent layers.
However, due to the DMI, the antiferromagnetic alignment is not exactly collinear, but there is a small canting in plane, in the same direction for all the spins, 
resulting in a net macroscopic magnetization.
The canting is a direct manifestation of the DMI, both in magnitude and in sign.
The magnitude of the canting angle or, equivalently, the ratio of the net magnetization to the sublattice 
magnetization is of the order of a few mrad \cite{Borovik1959,Borovik1961,Kreines,Petrov,Kobler} (Table \ref{TabDM}). 
Remarkably, it does not vary with the temperature below the onset of magnetic order \cite{Petrov}.

The relation between DMI and ferromagnetic moment can be grasped by assuming that the single-ion anisotropy allows the spins to rotate freely in the $ab$ plane, and writing the classical Hamiltonian for nearest-neighbor spins as:
\begin{equation}
 {\cal H}=J\bm{S}_1 \cdot \bm{S}_2+\bm{D} \cdot \left[ \bm{S}_1 \times \bm{S}_2 \right]
 \label{Hamiltonian}
\end{equation}
which minimizes energy by canting the spins with a small angle $\phi\sim \frac{1}{2}|\bm{D}|/|J|$.

{\it Diffraction experiment.} 
While the relative magnitude of the DMI is easy to determine from the canting angles, 
its sign has been reported only in our recent study of FeBO$_3$ \cite{DmitrienkoNP}.
In order to determine the sign $\sigma_\phi$ of the DMI, one needs to find the sign of the antiferromagnetic spin structure factor, which for (0,0,6n+3) reflections is simply the 
difference between the spin vectors at site 1 (Fig.~\ref{FIG1}) and one of its nearest neighbours; 
$\bm{S}_{AFM}=\bm{S}_1-\bm{S}_2$.
The macroscopic ferromagnetic moment can be aligned by a weak external magnetic field, which allows the entire magnetic structure to be rotated within the $ab$ plane.
While the intensity of magnetic scattering is easily determined, the all-important sign is 
lost when measuring the intensity of pure magnetic reflections with X-rays or neutrons.
We therefore exploit the interference between two X-ray amplitudes, one of magnetic origin, and a reference amplitude which is independent of the magnetic structure \cite{DmitrienkoNP}.
The former is dominated by X-ray non-resonant magnetic scattering \cite{deBergevin}, while the latter is quadruplole Resonant Elastic X-ray Scattering (REXS) \cite{GrenierREXS}. 
More details on both amplitudes are given in the Supplemental Material.
The interference is measured at the 009 Bragg reflection of the crystals of interest, which is forbidden for Thomson scattering (\textit{i.e.}~spacegroup forbidden) 
but allowed for the two scattering mechanisms outlined above.

It is, perhaps, worth noting that the sign of the DMI does not affect the direction of the ferromagnetic moment as it follows the external field. Rather, it 
determines whether one ferromagnetically-aligned sheet points to the left, and the one above it to the right, or vice versa. This difference is simply the phase of
the magnetic modulation, which is encoded in the phase of the magnetic scattering.

\begin{figure*}[!ht]
 \includegraphics[width=0.15\textwidth]{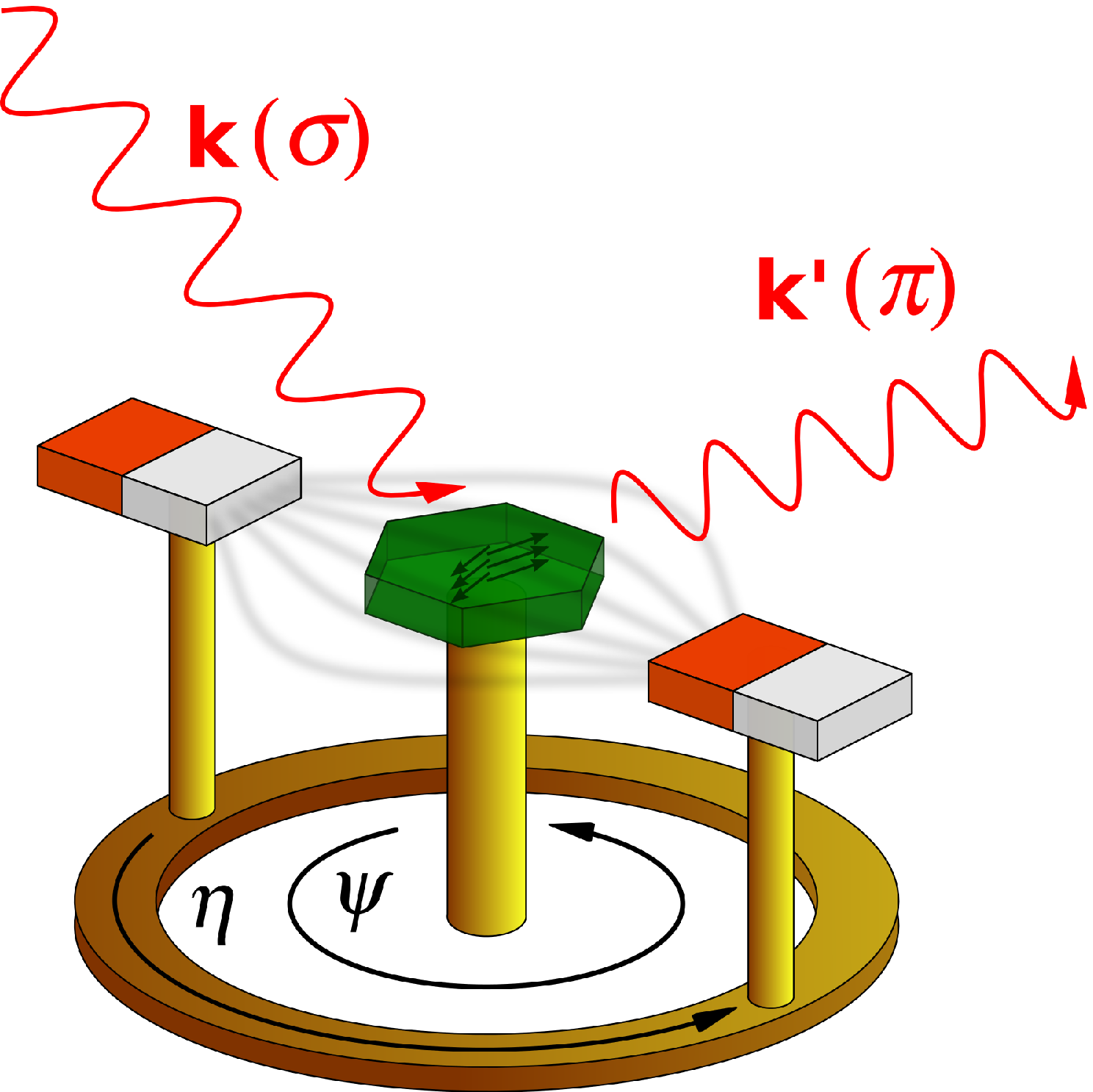}
 \includegraphics[width=0.2\textwidth]{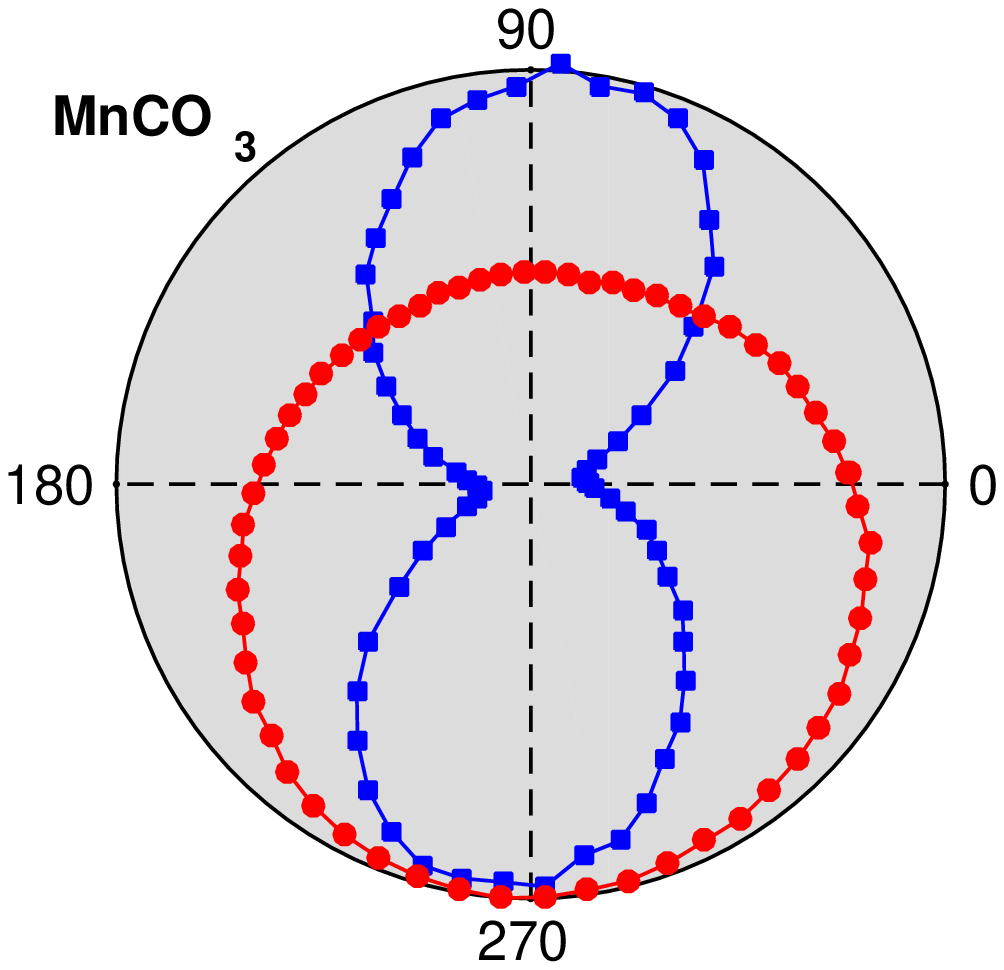}
 \includegraphics[width=0.2\textwidth]{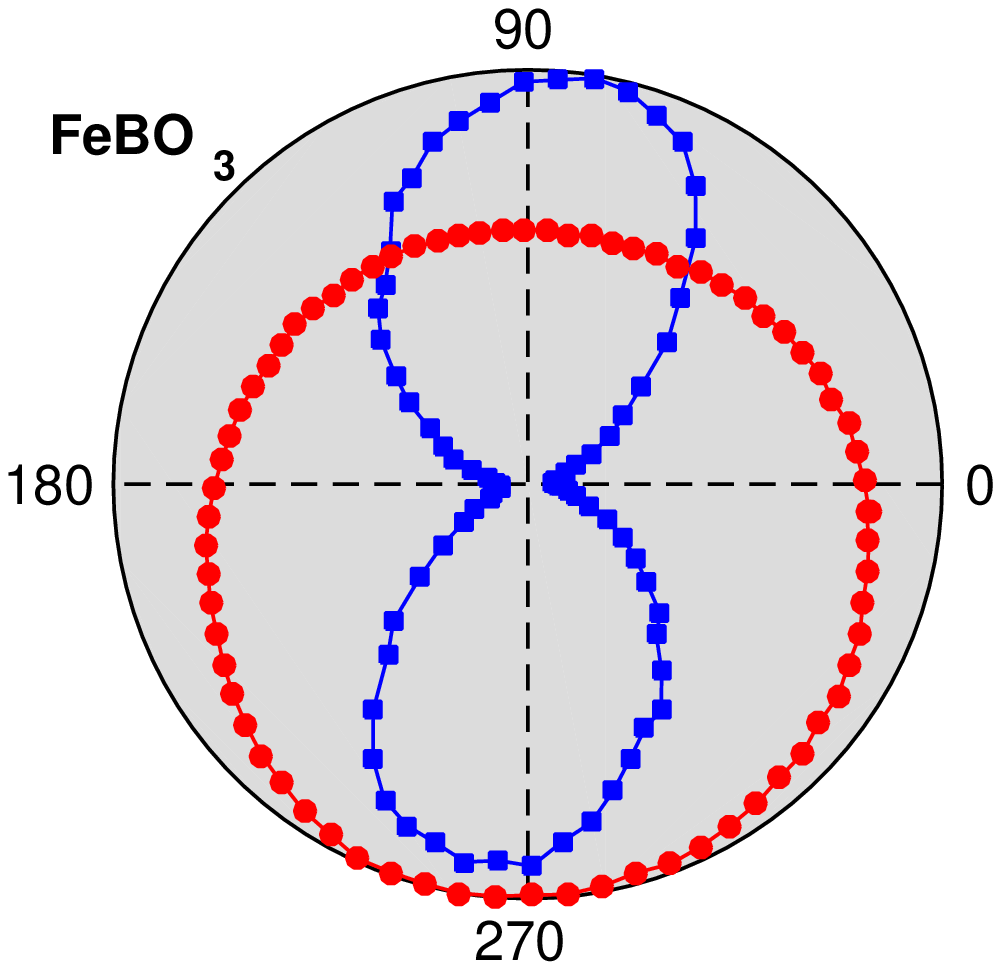}
 \includegraphics[width=0.2\textwidth]{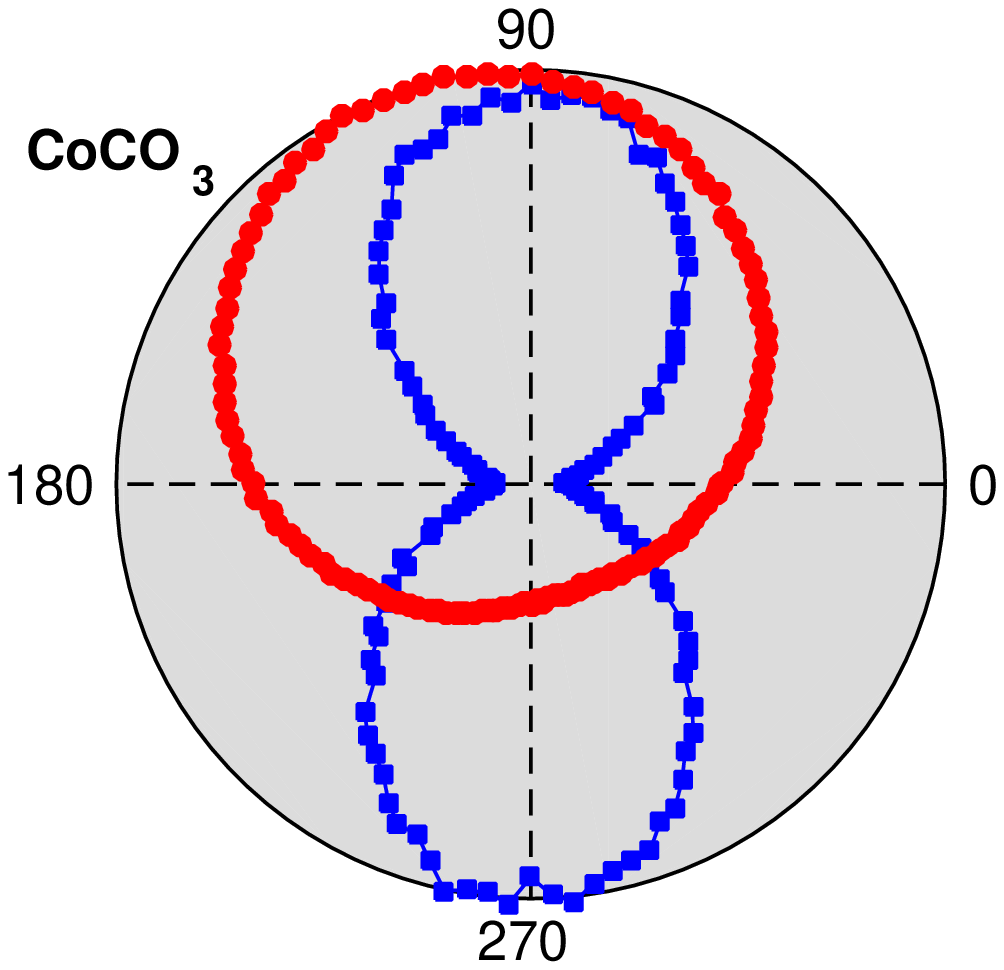}
 \includegraphics[width=0.2\textwidth]{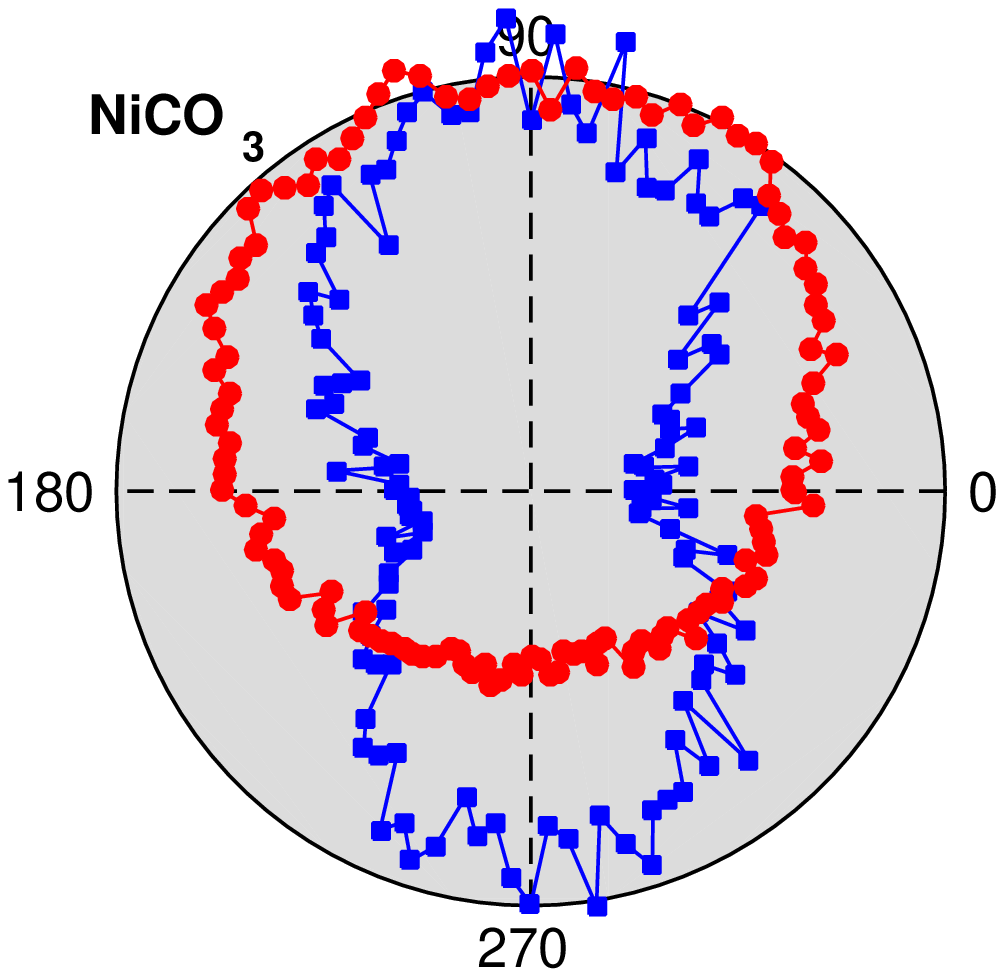}
 \caption{
 X-ray diffraction experiment: schematic view and main results.
 Normalized experimental values of the diffraction intensity $versus$ magnet angle $\eta$, for the series of weak ferromagnets. 
 The blue curves are measured below the resonance energy and show the pure magnetic scattering intensity, which is symmetric and insensitive to the scattering phase. 
 The red curves are on resonance and include a strong interference term that breaks the symmetry and gives the phase of the magnetic scattering, revealing the sign of the DMI.
}
 \label{FIG3}
\end{figure*}

The diffraction experiments reported here use the same set-up as that described in Ref.~\cite{DmitrienkoNP}.
We measured the 009 forbidden reflection of the four crystals with monochromatic X-rays tuned to the K-edge resonance of their respective magnetic ion.
The samples were macroscopic single crystals of high quality with a large 001 facet, except for the NiCO$_3$ crystal which was a grain of a few tens of microns.
The measurements were performed well below their respective N\'eel temperature, at 300 K, 7.5 K, 13 K and 5.5 K for, respectively, FeBO$_3$, MnCO$_3$, CoCO$_3$ and NiCO$_3$.
A $\sim$0.01 T magnetic field, sufficient to produce a single domain state aligned with the magnet, 
was applied by a pair of permanent magnets rotated about the sample $c$-axis by an angle $\eta$ (see Fig.~\ref{FIG3}).
The crystals were rotated by azimuthal angles $\psi$ about the scattering vector, to 
suitable orientations for the measurements (see Supplemental Material for details).
Measurements were performed at beamline I16 of Diamond Light Source \cite{I16}, using linearly polarised X-rays and a linear polarization analyser crystal to 
reject the scattered X-rays of unrotated polarisation. 

As a coherent sum of two scattering amplitudes, the diffraction intensity is the sum of a pure magnetic term, a pure resonant term, and an interference term
(see Supplemental Material):
\begin{widetext}
\begin{equation}
 I(E,\psi,\eta) = f_m^2 \sin^2\eta + \left|Q(E)\right|^2 \cos^2 3\psi + 2 \sigma_\phi f_m \Im\left[Q(E)\right] \cos3\psi \sin\eta
 \label{interf}
\end{equation}
\end{widetext}
where $f_m$ is a known positive quantity related to the non-resonant magnetic scattering amplitude, $E$ is the X-ray energy, and $Q(E)$ is a complex spectroscopic term related to the REXS amplitude. 
The latter can be calculated with a X-ray spectroscopy software such as FDMNES \cite{FDMNES}, which was used in this work.
From Eq.~(\ref{interf}), it is clear that one can extract the sign of the DMI ($\sigma_\phi$) by rotating the magnetic field while maintaining a fixed crystal 
azimuth ($\psi$) and X-ray energy ($E$). 
The results of such measurements are presented in Fig.~\ref{FIG3}.

\begin{table*}
\caption{Experimental and theoretical values of the canting angle (degrees). 
The experimental magnitudes are taken from the literature. 
The experimental signs and the \textit{ab initio} values are taken from this work.
The sign of the canting angle corresponds to the sign of the DMI. 
N$_{3d}$ is the number of the $3d$ electrons per transition metal site obtained from first-principles calculations (for details, see Supplemental Material).}
\label{TabDM}
\begin{tabular}{cccccccc}
  \hline
  \hline
  Compound	   & Magnetic & Z  & N$_{3d}$ & Canting angle $\phi$ (deg.) & Canting angle $\phi$ (deg.)\\
	   	   & ion      &    &          & experimental                &  \textit{ab initio} \\
  \hline
  MnCO$_3$ & Mn$^{2+}$ & 25 & 5.0 & -0.04 \cite{Kosterov2006234} , -0.4 \cite{Borovik1959,Kreines} & -0.05 \\
  FeBO$_3$ & Fe$^{3+}$ & 26 & 5.8 & -0.9 \cite{Petrov}                              & -0.8 \\
  CoCO$_3$ & Co$^{2+}$ & 27 & 7.1 & 4.9 \cite{Borovik1961,Kreines}   & 4.7 \\
  NiCO$_3$ & Ni$^{2+}$ & 28 & 8.2 & 10.8 \cite{Kreines}                           & 7.4 \\
\hline
\hline
\end{tabular}
\end{table*}

The sign of the magnetic structure factor is determined by the deviation of the measured intensity toward $\eta=90^\circ$ or $\eta=270^\circ$, {\it i.e.} whether 
the red rings in Fig.~\ref{FIG3} go up or down. The results are remarkably clear:
the sign of the DMI is the same in FeBO$_3$ and MnCO$_3$, which are both opposite to CoCO$_3$ and NiCO$_3$.
More precisely, 
the canting angle is negative (Fig.~\ref{FIG1}, left) in FeBO$_3$ and MnCO$_3$, and positive (Fig.~\ref{FIG1}, right) in CoCO$_3$ and NiCO$_3$.
These signs represent the missing information from the absolute values of the canting angles that are reported in the literature (Table~\ref{TabDM}) and complete our knowledge of the relative strength of the DMI in this series of materials.

A reliable model for the resonant spectrum 
$Q(E)$ (and in particular its imaginary part)
is a key requisite for the  correct interpretation of the scattering phase. A series of measurements at various energies and azimuths
confirmed the shape of the resonance, predicted by FDMNES, and showed that the resonant amplitude just below the resonance energy has a phase that is independent of the 
magnetic ion ($3d$ shell filling). This seemingly complex scattering process therefore provides a robust and consistent reference wave, and a reliable interpretation of the 
interference data.
At photon energies far below the core-level resonances 
the resonant term vanishes ($Q(E)\rightarrow 0$) and pure magnetic scattering is observed (the $\sin^2\eta$ term of Eq.~(\ref{interf})): the data become symmetric (Fig.~\ref{FIG3}), loosing all information about the scattering phase.

{\it First-principles calculations.} 
To simulate the electronic structure and magnetic properties of the selected compounds we used the Vienna {\it ab-initio} simulation package (VASP) \cite{VASP1993,VASP1996} 
within local density approximation taking into account the on-site Coulomb interaction $U$ and SOC (LDA+$U$+SO) \cite{Solovyev}. 
All the technical details are presented in the Supplemental Material. 

Table~\ref{TabDM} gives a summary of the main theoretical and experimental results.
One can see that the theory reproduces the change of the DMI sign through the series of studied compounds, observed experimentally. 
While the absolute values of the canting angles are slightly variable depending on the $U$ value used in the calculation, their signs turn out to be robust.  
Importantly, our first-principles calculations revealed that the chemical bonding in all four systems has more covalent rather than ionic character, as indicated
by the deviation of the number of the $3d$ electrons from the pure ionic values, and magnetization of the oxygen atoms. 

{\it Toy model.} 
According to Hund's rules, once the electronic shell becomes more than half-filled, the preferable mutual orientation of the spin and orbital moments changes.
It is tempting to use this argument to explain the change of the sign of the DMI across the series of carbonates. 
However, the present examples all have more-than-half-filled 3$d$ shells, and therefore parallel spin and orbital moments. We must therefore look further 
for an explanation of the microscopic mechanism behind the preferred magnetic chirality.

Here we propose a simple and transparent microscopic explanation of the DMI sign change in the $R\bar{3}c$ insulators, based on a minimal `toy' model.
The first step is to express the total Dzyaloshinskii-Moriya interaction between two atoms $i$ and $j$ as a sum of partial inter-orbital contributions (IO-DMI), $\mathbf{D}_{ij} = \sum_{nn'} \mathbf{D}^{nn'}_{ij}$. 
Here $n$ ($n'$) denote the half-filled states, which are magnetic and therefore contribute to the formation of the total spin moment of each atom.

Then we analyze the IO-DMI by means of a superexchange-based approach developed by Moriya \cite{Moriya2}. 
\begin{equation}
  \mathbf{D}^{nn'}_{ij} = \frac{8i}{U} [t^{nn'}_{ij} \mathbf{C}^{n'n}_{ji} - \mathbf{C}^{nn'}_{ij}t^{n'n}_{ji}],
\end{equation}
where $t^{nn'}_{ij}$ is the (unperturbed) hopping integral between n$^{th}$ ground orbital state of i$^{th}$ atom and n'$^{th}$ orbital state of j$^{th}$ atom , $\mathbf{C}^{nn'}_{ij}$ is the corresponding hopping renormalised by SOC and $U$ is the on-site Coulomb interaction. 
Since $t^{nn'}_{ij}=t^{n'n}_{ji}$ and the hoppings with SOC are imaginary, then the IO-DMI is non zero if $\mathbf{C}^{nn'}_{ij} = (\mathbf{C}^{n'n}_{ji})^{*}$.
This SOC-affected hopping integral is the quantity of interest, since it contains the information about the DMI sign.
As it was shown by Moriya \cite{Moriya2}, $\mathbf{C}^{nn'}_{ij}$ is related to the transfer of the electrons (holes) between the half-filled ground states and the excited ones. The latter can be either empty or fully occupied. 

\begin{figure*}
\centering
\includegraphics[width=0.8\textwidth]{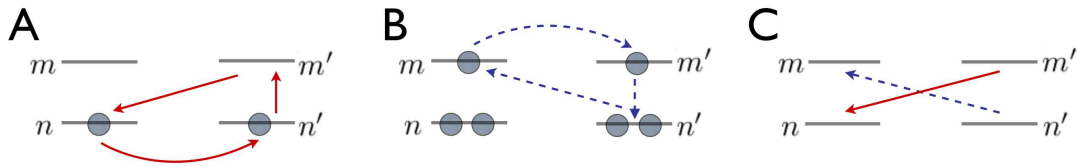}
\caption {Visualization of the toy tight-binding model proposed for explaining the change of DMI sign in the weak ferromagnets of this work. The filling of the energy levels in the ground state electronic configuration for N=2 ($\mathbf A$, left) and N=6 ($\mathbf B$ center) is shown. Arrows denote hopping (inter-atomic) and spin-orbit coupling (intra-atomic) excitations (between orbital states $n, n', m, m'$) corresponding to the Dzyaloshinskii-Moriya interaction. ($\mathbf C$, right) Comparison of the $\mathbf A$ and $\mathbf B$ excitation processes reveals the difference in hoppings between excited and ground states.}
\label{paths}
\end{figure*}

Importantly, the particular electronic configuration of the excited orbital states is related to the sign of the IO-DMI. 
To demonstrate this, we consider the simplest two-orbital two-site model with the different number of electrons, N=2 (Fig.~\ref{paths}~A)  and N=6 (Fig.~\ref{paths}~B). 
Here $n$ and $n'$ are the ground state orbitals in case A, while $m$ and $m'$ are the ground state orbitals in case B.
For simplicity, we assume that the hoppings between orbital states of the same symmetry are the same, $t_{ij}^{nn'} = t_{ij}^{mm'}$.  We fix the hopping integrals in our consideration, which means that the geometry of the model system does not change when we vary the number of electrons.

For our toy model we found that $\mathbf{C}_{21}^{n' \, n} = - \mathbf{C}_{21}^{m' \, m}$ (see Supplemental Material), simply because $\mathbf{C}_{21}^{n' \, n} \sim (t_{21}^{m'n} - t_{21}^{n'm})$  and  $\mathbf{C}_{21}^{m' \, m} \sim (t_{21}^{n'm} - t_{21}^{m'n})$ (Fig.~\ref{paths} $\mathbf C$).
It means that $\mathbf{D}^{nn'}_{12}$ (N=2) = - $\mathbf{D}^{mm'}_{12}$ (N=6). The change of DMI sign is robust, even if $t^{nn'}_{12} \ne t_{12}^{mm'}$. Thus, already at the level of the Moriya's approach, the sign of the DMI is shown to depend on the orbitals filling. 
By changing the occupation of the $3d$ states we change the balance between empty and fully occupied channels for the IO-DMI.
It results in the change of sign of the total DMI. In contrast to the previous considerations on metals \cite{PhysRevLett.44.1538,kashid,Belabbes} with complex dependence of the DMI energy on the electronic structure, our toy model for insulators puts forward an intuitive picture of DMI. 

To summarize, 
we have performed a systematic experimental and theoretical investigation of the Dzyaloshinskii-Moriya antisymmetric exchange interaction in a series of isostructural weak ferromagnets, 
and have discovered and explained a dramatic
variation in magnitude and sign as the $3d$ orbitals are gradually filled. Our novel x-ray diffraction technique yields both the amplitude and phase of the magnetic diffraction, essential for 
determining its sign. We have shown that it is suitable even for very small (few tens of microns) crystal samples. The dramatic evolution of the Dzyaloshinskii-Moriya interaction
with electron filling, and the ability of 
modern first-principles calculations to model it, bodes very well for a future in which the exchange interactions can be tuned for spintronics technologies, and important materials 
properties predicted by computational methods.

\section{Acknowledgement} 
The authors acknowledge Diamond Light Source for time on Beamline I16 under Proposals MT-7703 and MT-11751 and the XMaS facility at ESRF for beamtime under proposal BM-28-01-966.
We acknowledge fruitful communications with P. J. Brown, N. M. Kreines, Igor Solovyev, Alexander Tsirlin and Frederic Mila, and Yves Joly for his valuable help with the FDMNES code. 
We are in debt of N. M. Kreines for providing the MnCO$_3$ and CoCO$_3$ single crystals.
The work of V.V.M. is supported by the grant of the President of Russian Federation MD-6458.2016.2.
A.I.L. acknowledges the support of DFG  SFB-668 and the excellence cluster CUI. 
M.I.K. acknowledges support from the European Union, Horizon 2020 research and innovation programme under grant agreement No. 696656, GrapheneCore1.
Y.O.K. acknowledges the computational resources provided by the Swedish National Infrastructure for Computing (SNIC) and Uppsala Multidisciplinary Center for Advanced Computational Science (UPPMAX).
The list of authors follows the alphabetical order and is not indicative of the importance of the contribution of each.


%

\end{document}